\documentclass[aps,prl,twocolumn,superscriptaddress,showpacs,preprintnumbers,amsmath,amssymb,floatfix,nofootinbib]{revtex4}

\usepackage{graphicx}
\usepackage{dcolumn}
\usepackage{bm}
\usepackage{url} 
\usepackage{amsmath} 
\usepackage{amssymb} 

\begin{document}
\title{
Observation of $D^0-\bar{D}^0$ Mixing in $e^+e^-$ Collisions
}
\noaffiliation
\affiliation{University of the Basque Country UPV/EHU, 48080 Bilbao}
\affiliation{Beihang University, Beijing 100191}
\affiliation{University of Bonn, 53115 Bonn}
\affiliation{Budker Institute of Nuclear Physics SB RAS and Novosibirsk State University, Novosibirsk 630090}
\affiliation{Faculty of Mathematics and Physics, Charles University, 121 16 Prague}
\affiliation{University of Cincinnati, Cincinnati, Ohio 45221}
\affiliation{Deutsches Elektronen--Synchrotron, 22607 Hamburg}
\affiliation{Justus-Liebig-Universit\"at Gie\ss{}en, 35392 Gie\ss{}en}
\affiliation{Gyeongsang National University, Chinju 660-701}
\affiliation{Hanyang University, Seoul 133-791}
\affiliation{University of Hawaii, Honolulu, Hawaii 96822}
\affiliation{High Energy Accelerator Research Organization (KEK), Tsukuba 305-0801}
\affiliation{Hiroshima Institute of Technology, Hiroshima 731-5193}
\affiliation{IKERBASQUE, Basque Foundation for Science, 48011 Bilbao}
\affiliation{Indian Institute of Technology Guwahati, Assam 781039}
\affiliation{Indian Institute of Technology Madras, Chennai 600036}
\affiliation{Institute of High Energy Physics, Chinese Academy of Sciences, Beijing 100049}
\affiliation{Institute of High Energy Physics, Vienna 1050}
\affiliation{Institute for High Energy Physics, Protvino 142281}
\affiliation{INFN - Sezione di Torino, 10125 Torino}
\affiliation{Institute for Theoretical and Experimental Physics, Moscow 117218}
\affiliation{J. Stefan Institute, 1000 Ljubljana}
\affiliation{Kanagawa University, Yokohama 221-8686}
\affiliation{Institut f\"ur Experimentelle Kernphysik, Karlsruher Institut f\"ur Technologie, 76131 Karlsruhe}
\affiliation{Kavli Institute for the Physics and Mathematics of the Universe (WPI), University of Tokyo, Kashiwa 277-8583}
\affiliation{Korea Institute of Science and Technology Information, Daejeon 305-806}
\affiliation{Korea University, Seoul 136-713}
\affiliation{Kyungpook National University, Daegu 702-701}
\affiliation{\'Ecole Polytechnique F\'ed\'erale de Lausanne (EPFL), Lausanne 1015}
\affiliation{Faculty of Mathematics and Physics, University of Ljubljana, 1000 Ljubljana}
\affiliation{University of Maribor, 2000 Maribor}
\affiliation{Max-Planck-Institut f\"ur Physik, 80805 M\"unchen}
\affiliation{School of Physics, University of Melbourne, Victoria 3010}
\affiliation{Moscow Physical Engineering Institute, Moscow 115409}
\affiliation{Graduate School of Science, Nagoya University, Nagoya 464-8602}
\affiliation{Kobayashi-Maskawa Institute, Nagoya University, Nagoya 464-8602}
\affiliation{Nara Women's University, Nara 630-8506}
\affiliation{National Central University, Chung-li 32054}
\affiliation{National United University, Miao Li 36003}
\affiliation{Department of Physics, National Taiwan University, Taipei 10617}
\affiliation{H. Niewodniczanski Institute of Nuclear Physics, Krakow 31-342}
\affiliation{Nippon Dental University, Niigata 951-8580}
\affiliation{Niigata University, Niigata 950-2181}
\affiliation{University of Nova Gorica, 5000 Nova Gorica}
\affiliation{Osaka City University, Osaka 558-8585}
\affiliation{Pacific Northwest National Laboratory, Richland, Washington 99352}
\affiliation{Panjab University, Chandigarh 160014}
\affiliation{Peking University, Beijing 100871}
\affiliation{University of Pittsburgh, Pittsburgh, Pennsylvania 15260}
\affiliation{RIKEN BNL Research Center, Upton, New York 11973}
\affiliation{University of Science and Technology of China, Hefei 230026}
\affiliation{Seoul National University, Seoul 151-742}
\affiliation{Soongsil University, Seoul 156-743}
\affiliation{Sungkyunkwan University, Suwon 440-746}
\affiliation{School of Physics, University of Sydney, NSW 2006}
\affiliation{Tata Institute of Fundamental Research, Mumbai 400005}
\affiliation{Excellence Cluster Universe, Technische Universit\"at M\"unchen, 85748 Garching}
\affiliation{Toho University, Funabashi 274-8510}
\affiliation{Tohoku Gakuin University, Tagajo 985-8537}
\affiliation{Tohoku University, Sendai 980-8578}
\affiliation{Department of Physics, University of Tokyo, Tokyo 113-0033}
\affiliation{Tokyo Institute of Technology, Tokyo 152-8550}
\affiliation{Tokyo Metropolitan University, Tokyo 192-0397}
\affiliation{Tokyo University of Agriculture and Technology, Tokyo 184-8588}
\affiliation{University of Torino, 10124 Torino}
\affiliation{CNP, Virginia Polytechnic Institute and State University, Blacksburg, Virginia 24061}
\affiliation{Wayne State University, Detroit, Michigan 48202}
\affiliation{Yonsei University, Seoul 120-749}
  \author{B.~R.~Ko}\affiliation{Korea University, Seoul 136-713} 
  \author{E.~Won}\affiliation{Korea University, Seoul 136-713} 
  \author{I.~Adachi}\affiliation{High Energy Accelerator Research Organization (KEK), Tsukuba 305-0801} 
  \author{H.~Aihara}\affiliation{Department of Physics, University of Tokyo, Tokyo 113-0033} 
  \author{K.~Arinstein}\affiliation{Budker Institute of Nuclear Physics SB RAS and Novosibirsk State University, Novosibirsk 630090} 
  \author{D.~M.~Asner}\affiliation{Pacific Northwest National Laboratory, Richland, Washington 99352} 
  \author{V.~Aulchenko}\affiliation{Budker Institute of Nuclear Physics SB RAS and Novosibirsk State University, Novosibirsk 630090} 
  \author{T.~Aushev}\affiliation{Institute for Theoretical and Experimental Physics, Moscow 117218} 
  \author{A.~Bala}\affiliation{Panjab University, Chandigarh 160014} 
  \author{V.~Bhardwaj}\affiliation{Nara Women's University, Nara 630-8506} 
% ...
\collaboration{The Belle Collaboration}
  
\begin{abstract}
We observe $D^0-\bar{D}^0$ mixing in the decay $D^0\rightarrow
K^+\pi^-$ using a data sample of integrated luminosity 976
fb$^{-1}$ collected with the Belle detector at the KEKB $e^+e^-$
asymmetric-energy collider. We measure the mixing parameters 
${x'}^2 = (0.09\pm0.22)\times 10^{-3}$ and 
$y' = (4.6\pm3.4)\times 10^{-3}$
and the ratio of doubly Cabibbo-suppressed to
Cabibbo-favored decay rates $R_D = (3.53\pm0.13)\times 10^{-3}$, 
where the uncertainties are statistical
and systematic combined. Our measurement excludes the no-mixing
hypothesis at the 5.1 standard deviation level. 
\end{abstract}
\maketitle

{\renewcommand{\thefootnote}{\fnsymbol{footnote}}}
\setcounter{footnote}{0}
A weakly decaying flavored neutral meson is a two-state quantum system
with an allowed transition between the two states.  This transition is
referred to as neutral meson mixing and originates from the difference
between the flavor and mass eigenstates of the meson-antimeson system
with a well-known rate depending on elements of the
Cabibbo-Kobayashi-Maskawa matrix~\cite{CABIBBO,KM}. Mixing phenomena
are well established for $K^0$, $B^0$, and $B^0_s$ mesons and their
mixing rates are consistent with predictions based on the standard
model (SM)~\cite{PDG2012}. $D^0$ mixing has also recently been
observed in hadron collider experiments~\cite{LHCBMIX, CDFMIX_PRE},
confirming a previous $D^0-\bar{D}^0$ mixing signal~\cite{HFAG} based
mainly on combined evidence from three different
experiments~\cite{BABARMIX, BELLEMIX, CDFMIX_OLD}.

The phenomenology of meson mixing is described by two
parameters, $x=\Delta m/\Gamma$ and $y=\Delta\Gamma/2\Gamma$, where
$\Delta m$ and $\Delta\Gamma$ are the mass and width differences between
the two mass eigenstates and $\Gamma$ is the average decay width of
the mass eigenstates. 
While the finite mixing parameters of the $K^0$, $B^0$, and $B^0_s$
mesons are well measured, those for the $D^0$ meson are
not~\cite{HFAG}.
The mixing parameters $x$ and $y$ are difficult to
calculate~\cite{BIANCO, FALK}, which complicates the interpretation of
experimental measurements against the SM. Nevertheless, it is still of
great interest to improve the measurement of the $D^0$ mixing
parameters to search for possible beyond-SM physics
contributions~\cite{BSM}. It is also very valuable to confirm $D^0$
mixing in $e^+e^-$ collisions and provide further independent
determinations of the $D^0$ mixing parameters where the experimental
conditions are quite different from those in hadron collider
experiments.

In this Letter, we report the first observation of $D^0-\bar{D}^0$
mixing from an $e^+e^-$ collision experiment by measuring the
time-dependent ratio of the $D^0\to K^+\pi^-$ to $D^0\to
K^-\pi^+$ decay rates.  The consideration of charge-conjugated
decays is implied throughout this Letter.
We refer to $D^0\to K^+\pi^-$ as wrong-sign (WS)
and $D^0\to K^-\pi^+$ as right-sign (RS) decays.
We tag the RS and WS decays through the decay chain
$D^{*+}\to D^0(\to K^{\mp}\pi^{\pm})\pi^+_s$ by comparing the charge
of the $\pi$ from the $D^0$ decay and the charge of the low-momentum
$\pi_s$ from the $D^{*+}$ decay. The RS decay amplitude is the sum of
the amplitudes for Cabibbo-favored (CF) decay $D^0\to K^-\pi^+$ and
$D^0-\bar{D}^0$ mixing followed by the doubly-Cabibbo-suppressed (DCS)
decay $\bar{D}^0\to K^-\pi^+$, where the latter is very small compared
to the former and is therefore neglected.
In contrast, the WS decay amplitude is the sum of two comparable decay
amplitudes for the DCS decay $D^0\to K^+\pi^-$ and
$D^0-\bar{D}^0$ mixing followed by the CF decay $\bar{D}^0\to
K^+\pi^-$. Assuming charge-conjugation and parity ($CP$)
conservation and that the mixing parameters are small ($|x|\ll 1$ and
$|y|\ll 1$), the time-dependent RS and WS decay rates are
\begin{eqnarray}
  \nonumber
  \Gamma_{\rm RS}(\tilde{t}/\tau)&\approx&|\mathcal{A}_{\rm CF}|^2 e^{-\frac{\tilde{t}}{\tau}},\\
  \nonumber
  \Gamma_{\rm WS}(\tilde{t}/\tau)&\approx&|\mathcal{A}_{\rm CF}|^2 e^{-\frac{\tilde{t}}{\tau}}\\
  &\times&\Biggl(R_D + \sqrt{R_D}y'\frac{\tilde{t}}{\tau} + \frac{{x'}^2+y'^2}{4}\biggl(\frac{\tilde{t}}{\tau}\biggr)^2\Biggr)
  \label{EQ:GAMMA_THEORY}
\end{eqnarray}
to second order in the mixing parameters. In
Eq.~(\ref{EQ:GAMMA_THEORY}), $\tilde{t}$ is the true proper decay
time, $\mathcal{A}_{\rm CF}$ is the CF decay amplitude, $\tau$ is the
$D^0$ lifetime, $R_D$ is the ratio of DCS to CF decay rates,
$x'=x\cos\delta+y\sin\delta$, and $y'=y\cos\delta-x\sin\delta$, where
$\delta$ is the strong phase difference between the DCS and CF decay
amplitudes. The time-dependent ratio of WS to RS decay rates is then
\begin{equation}
  R(\tilde{t}/\tau)=\frac{\Gamma_{\rm WS}(\tilde{t}/\tau)}{\Gamma_{\rm RS}(\tilde{t}/\tau)}
  \approx R_D + \sqrt{R_D}y'\frac{\tilde{t}}{\tau} + \frac{{x'}^2+y'^2}{4}\biggl(\frac{\tilde{t}}{\tau}\biggr)^2,
  \label{EQ:RWS_THEORY}
\end{equation}
which is a quadratic function of $\tilde{t}/\tau$.

\begin{figure}[htbp]
  \includegraphics[height=0.35\textwidth,width=0.5\textwidth]{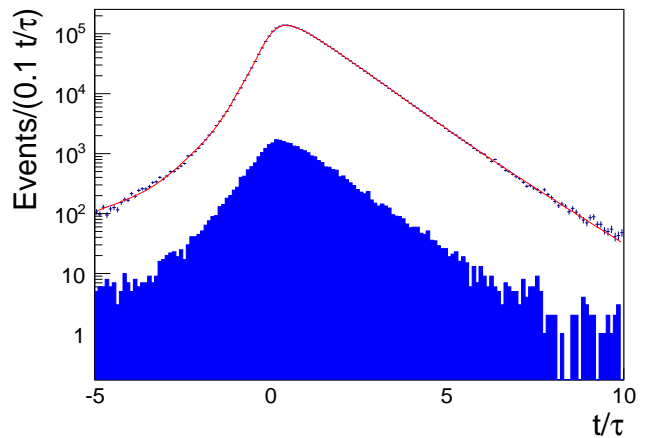}
  \caption{Distribution of the proper decay time from
    background-subtracted RS decays in the signal region (points with
    error bars) and in the sideband regions (shaded). The curve shows
    the fit to the signal.}
  \label{FIG:DTIME}
\end{figure}

In order to measure the mixing parameters using
Eq.~(\ref{EQ:RWS_THEORY}), the measured proper decay time should be
approximately the true proper decay time. 
This condition is satisfied in hadron collider
experiments~\cite{LHCBMIX, CDFMIX_PRE} where the tagged $D$'s have a
decay time much larger than the resolution on $\tilde{t}$. At a
$B$-factory, however, the mean decay time of the tagged $D$'s, shown
in Fig.~\ref{FIG:DTIME}, is approximately the $D^0$ lifetime, which is
comparable to the resolution on $\tilde{t}$; thus, the resolution
effect must be taken into account. Our approach here is to measure the
time-dependent ratio of WS to RS decays, given by
\begin{equation}
  R(t/\tau)=\frac{\int^{+\infty}_{-\infty}\Gamma_{\rm WS}(\tilde{t}/\tau)\mathcal{R}(t/\tau-\tilde{t}/\tau)d(\tilde{t}/\tau)}{\int^{+\infty}_{-\infty}\Gamma_{\rm RS}(\tilde{t}/\tau)\mathcal{R}(t/\tau-\tilde{t}/\tau)d(\tilde{t}/\tau)},
  \label{EQ:RWS_EXP}
\end{equation}
where $t$ is the reconstructed proper decay time and
$\mathcal{R}(t/\tau-\tilde{t}/\tau)$ is the resolution function of the
real decay time, $\tilde{t}$.  

The data used in this analysis are recorded at the $\Upsilon(nS)$
resonances $(n=1,2,3,4,5)$ or near the $\Upsilon(4S)$ resonance with
the Belle detector at the $e^+e^-$ asymmetric-energy collider
KEKB~\cite{KEKB}. The data sample corresponds to an integrated
luminosity of 976 fb$^{-1}$. The Belle detector is a large-solid-angle
magnetic spectrometer that consists of a silicon vertex detector
(SVD), a 50-layer central drift chamber (CDC), an array of aerogel
threshold Cherenkov counters (ACC), a barrel-like arrangement of
time-of-flight scintillation counters (TOF), and an electromagnetic
calorimeter comprising CsI(Tl) crystals (ECL) located inside a
superconducting solenoid coil that provides a 1.5 T magnetic field. An
iron flux return located outside the coil is instrumented to detect
$K^0_L$ mesons and identify muons. A detailed description of the
Belle detector can be found in Ref.~\cite{BELLE}.

We require that charged tracks originate from the $e^+e^-$ interaction
point (IP) with an impact parameter less than 4 cm in the beam
direction (the $z$ axis) and 2 cm in the transverse plane and have a
transverse momentum greater than 0.1 GeV$/c$. All
charged tracks are required to have at least two associated hits each
in the $z$ and azimuthal strips of the SVD to assure good spatial
resolution of the decay vertices of $D^0$ mesons.
Charged tracks are identified as $K$ or $\pi$ candidates using the ratio of
particle identification likelihoods,
$\mathcal{P}_{K\pi}\equiv\mathcal{L}_K/(\mathcal{L}_K+\mathcal{L}_\pi)$,
reconstructed from the track-associated data in the CDC, TOF, and ACC. We require
$\mathcal{P}_{K\pi}>0.4$ for $K$, $\mathcal{P}_{K\pi}<0.7$ for $\pi$
and $\mathcal{P}_{K\pi}<0.9$ for $\pi_s$ candidates. The efficiency
and $K/\pi$ misidentification rate of the $K$ selection are 91\% and
12\% and those of the $\pi$ selection are 94\% and 18\%. We also apply a
loose electron veto criterion using the ECL information for all charged
tracks.  Oppositely-charged $K$ and $\pi$ candidates are combined to form a $D^0$
candidate by fitting them to a common vertex; the resulting $D^0$ candidate
is fit to the IP to give the $D^{*+}$ vertex. A $D^{*+}$ candidate
is reconstructed by combining a $D^0$ candidate---a $K\pi$ combination
with invariant mass within $\pm$20 MeV/$c^2$ (i.e.,
$\sim$$\pm3\sigma$) of the nominal $D^0$ mass~\cite{PDG2012}---with a
$\pi_s$. The $\pi_s$ is further constrained to pass through the
$D^{*+}$ vertex. The sum of the reduced $\chi^2$ of the $D^{*+}$
vertex fit and $\pi_s$ fit to the $D^{*+}$ vertex is required to be
less than 16. 

There is a significant contribution to the WS sample from RS decays
where both $K$ and $\pi$ candidates are misidentified as $\pi$ and
$K$, respectively. We remove these with tighter particle
identification requirements,
$\mathcal{P}_{K\pi}>0.99$ for $K$ and
$\mathcal{P}_{K\pi}<0.01$ for $\pi$, if $M(K\pi)_{\rm swap}$, the
invariant mass of the $K\pi$ combination when swapping the nominal mass of
$K$ and $\pi$ track candidates, is within $\pm$25 MeV/$c^2$ of the
nominal $D^0$ mass. 
To remove combinatorial background due to random unassociated charged track
combinations that meet all the other requirements, we require the
$D^{*+}$ meson momentum calculated in the center-of-mass system to be
greater than 2.5, 2.6, and 3.0 GeV/$c$ for the data taken below the
$\Upsilon(4S)$, at the $\Upsilon(4S)$, and above the $\Upsilon(4S)$
resonance, respectively. This momentum requirement also removes
$D^{*+}\to D^0\pi^+_s$ decays from $B$ meson decays, which do not give
the proper decay time of the $D^0$ meson due to the finite $B$-meson
lifetime.

The selection criteria described above are chosen by maximizing $R_{\rm
  WS}\mathcal{N}^{\rm RS}_{\rm S}/\sqrt{R_{\rm WS}\mathcal{N}^{\rm
    RS}_{\rm S}+\mathcal{N}^{\rm WS}_{\rm B}}$, where $R_{\rm WS}$ is
the nominal ratio of WS to RS decay rates~\cite{PDG2012},
$\mathcal{N}^{\rm RS}_{\rm S}$ is the number of events in the RS
signal region of the $D^{*+}$-$D^0$ mass difference, $\Delta M\equiv
M(D^{*+}\to D^0(\to K\pi)\pi^+_s)-M(D^0\to K\pi)$, and
$\mathcal{N}^{\rm WS}_{\rm B}$ is that in the WS sideband regions of
$\Delta M$.
We define the signal region as $\Delta M\in[0.144,0.147]$ GeV/$c^2$
and the background sidebands as $\Delta M\in[0.141,0.142]$ or
$[0.149,0.151]$ GeV/$c^2$.  
When counting $\mathcal{N}^{\rm RS}_{\rm S}$, we subtract background
candidates in the signal region using candidates in the RS sideband
regions.

The measured $D^0$ proper decay time is calculated as
$t=m_{D^0}\vec{L}\cdot\vec{p}/|\vec{p}|^2$ where $\vec{L}$ is the
vector joining the decay and production vertices of the $D^0$, $\vec{p}$
is the $D^0$ momentum, and $m_{D^0}$ and $\tau$ are the nominal $D^0$
mass and lifetime~\cite{PDG2012}.
We require the uncertainty on $t$ to satisfy $\sigma_t/\tau<1.0$, and
$t/\tau\in[-5,10]$. These selections are determined  from 5000
simplified simulated experiments by maximizing our sensitivity to the
mixing parameters and minimizing the systematic biases in them.

\begin{figure}[htbp]
  \includegraphics[height=0.23\textwidth,width=0.47\textwidth]{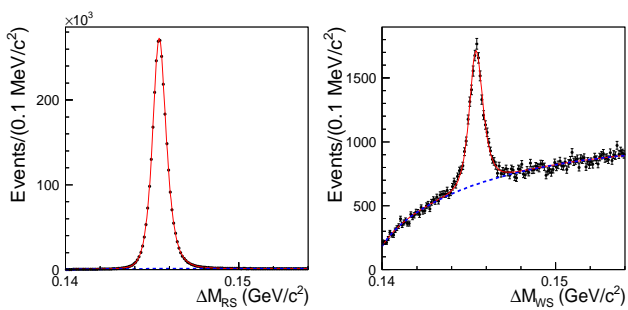}
  \caption{Time-integrated distributions for the mass difference of RS
    (left) and WS (right) candidates. Points with error bars are the
    data; full and dashed lines are, respectively, the signal and
    background fits described in the text.}
\label{FIG:DELM}
\end{figure}

Using these selections, we find no significant backgrounds in WS
candidates that peak in the signal region from a large-statistics
sample of fully simulated $e^+e^-\to$ hadrons events in our
GEANT3-based \cite{MC} Monte Carlo (MC)
simulation. Figure~\ref{FIG:DELM} shows the time-integrated
distributions of $\Delta M$ from RS and WS candidate events after
applying all the selections described above.

The time-integrated RS signal shown in Fig.~\ref{FIG:DELM} is
parametrized as a sum of Gaussian and Johnson $S_U$~\cite{JOHNSON}
distributions with a common mean. The time-dependent RS signal in each
bin of the proper decay time is fit with a Johnson $S_U$ only. The
shapes of the WS signal are fixed using the corresponding RS signal
shapes, and fit with only the signal normalization allowed to
vary. The backgrounds in RS and WS decay events are fit independently
and are parametrized with the form $(\Delta M-m_{\pi^+})^\alpha
e^{-\beta(\Delta M-m_{\pi^+})}$, where $\alpha$ and $\beta$ are free
fit parameters, and $m_{\pi^+}$ is the nominal mass of
$\pi^+$~\cite{PDG2012}. The fits give 2~980~710$\pm$1885 RS and
11~478$\pm$177 WS decays, giving an inclusive ratio of WS to RS decay
rates of $(3.851\pm0.059)\times10^{-3}$. The uncertainty is
statistical only.

We obtain the resolution function of Eq.~(\ref{EQ:RWS_EXP}) from the
proper decay time distribution of RS decays after subtracting a small
level of background events using the sideband regions defined
above. This is shown in Fig.~\ref{FIG:DTIME}.
We parametrize the proper decay time distribution of RS decays with
the convolution of an exponential and a resolution function that is
constructed as the sum of four Gaussians,
$\mathcal{R}(t/\tau)=\sum^4_{i=1}f_i G_i(t/\tau;\mu_i,\sigma_i)$,
where $G_i$ is a Gaussian distribution with mean $\mu_i$ and width
$\sigma_i$ and $f_i$ is its weight. The mean $\mu_i$
is further parametrized with $\mu_i=\mu_1+a\sigma_i$, where $\mu_1$ is
the mean of the core Gaussian $G_1$ $(i=2,3,4)$. The parameters $a$
and $\mu_1$ describe a possible asymmetry of the resolution
function. All parameters of the resolution function float freely
and the fit is shown in Fig.~\ref{FIG:DTIME}. The $D^0$ lifetime is also a
free fit parameter, for which we obtain $(408.5\pm0.9)$ fs,
where the uncertainty is statistical only. This $D^0$ lifetime is
consistent with the world-average value~\cite{PDG2012} and the other
Belle measurement~\cite{MARKO_CHARM12}, which gives further confidence
in our parametrization of the resolution function.

\begin{figure}[htbp]
  \includegraphics[height=0.35\textwidth,width=0.47\textwidth]{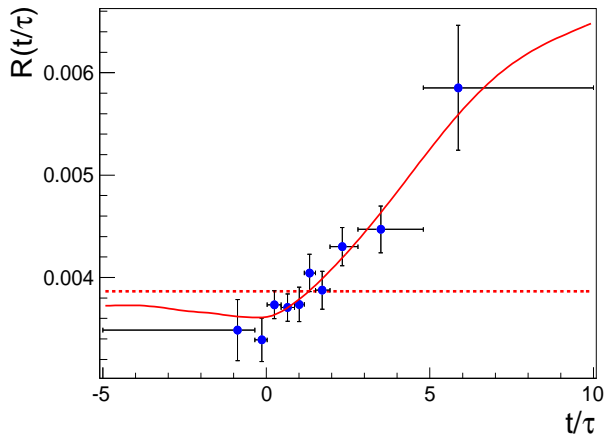}
  \caption{The time-dependent ratios of WS to RS decay rates. Points
    with error bars reflect the data and their total
    uncertainties. The lines show the fit with (solid) and without
    (dashed) the mixing hypothesis.}
\label{FIG:MIXING}
\end{figure}

To calculate the time-dependent WS to RS decay rate ratio, we divide
the samples shown in Fig.~\ref{FIG:DELM} into ten bins of proper decay
time. Our binning choice is made using 5000
simplified simulated experiments to maximize the sensitivity to the
mixing parameters. Figure~\ref{FIG:MIXING} shows the
time-dependent ratios of WS to RS decay rates. The average value of
the proper decay time in each bin is determined with the
parametrization for the reconstructed RS proper decay time
distribution shown in Fig.~\ref{FIG:DTIME}.

Prior to our fit to the time-dependent ratios of WS to RS decay
rates, we estimate possible systematic effects.  We validate the analysis
procedure with the fully simulated MC events with several
different input values of the mixing parameters and find results
consistent with the input parameters.
The dominant sources of systematic uncertainties are from 
fitting the $\Delta M$ distributions
and uncertainties on the resolution function that do not cancel out
in Eq.~(\ref{EQ:RWS_EXP}). However, these are estimated to be less
than a tenth of the statistical uncertainty, which is estimated in
simulated simplified experiments. Other sources of uncertainty are the
binning of the proper decay time and the reconstruction efficiencies
of WS and RS decays. These effects should cancel in the WS to RS ratio
measurement. We estimate these with simulated simplified experiments
and, indeed, find a negligible contribution of
$<$$\mathcal{O}(10^{-4})$ on the mixing parameters and so ignore
them. The systematic uncertainties due to fitting the $\Delta M$
distributions are estimated in the bins of the proper decay time and
are added to the statistical uncertainties of the bin in quadrature,
albeit with negligible effect.

Our fits to the time-dependent ratios of WS to RS decays using
Eq.~(\ref{EQ:RWS_EXP}) are shown in Fig.~\ref{FIG:MIXING}. We test two
hypotheses, with and without mixing, and the results are listed in
Table~\ref{TABLE:MIXRESULTS}.
\begin{table}[htbp]
\begin{ruledtabular}
\begin{center}
\caption{Results of the time-dependent fit to $R(t/\tau)$, where DOF
  stands for the degrees of freedom. The uncertainties are statistical
  and systematic combined.}
\label{TABLE:MIXRESULTS}
\begin{tabular}{cccccc} 
Test           &          &Fit              &&Correlation& \\
hypothesis     &Parameters&results          &&coefficient& \\
($\chi^2$/DOF) &          &($10^{-3}$)   &$R_D$&$y'$&${x'}^2$\\ \hline
Mixing         &$R_D$     &$3.53\pm0.13$ &1    &$-$0.865&$+$0.737\\
(4.2/7)        &$y'$      &$4.6\pm3.4$   &     &1     &$-$0.948\\ 
               &${x'}^2$    &$0.09\pm0.22$ &     &      &1\\ \hline
No Mixing      &$R_D$     &$3.864\pm0.059$&&&\\ 
(33.5/9)       &&&&&\\
\end{tabular}     
\end{center}
\end{ruledtabular}
\end{table}
The mixing parameters measured in this analysis agree with previous
results from both hadron collider experiments~\cite{CDFMIX_PRE,
  LHCBMIX2} using a similar method, as well as with the results of
alternate experimental methods from $e^+e^-$ collision
experiments~\cite{BABARMIX, OLDWS} and are summarized in
Table~\ref{TABLE:COMP}.
\begin{table}[htbp]
\begin{ruledtabular}
\begin{center}
\caption{Measured $D^0-\bar{D}^0$ mixing parameters in $D^0\rightarrow
  K^+\pi^-$ decays from this work and others, where we display the
total uncertainty. All measurements assume $CP$ conservation.}
\label{TABLE:COMP}
\begin{tabular}{cccc} 
Experiment             &$R_D$ ($\times10^{-3}$)&$y'$ ($\times10^{-3}$)&${x'}^2$ ($\times10^{-3}$)\\ \hline
Belle~\cite{OLDWS}     &$3.64\pm0.17$          &$0.6^{+4.0}_{-3 .9}$  &$0.18^{+0.21}_{-0.23}$  \\
BaBar~\cite{BABARMIX}  &$3.03\pm0.19$          &$9.7\pm5.4$           &$-0.22\pm0.37$          \\ 
CDF  ~\cite{CDFMIX_PRE}&$3.51\pm0.35$          &$4.3\pm4.3$           &$0.08\pm0.18$           \\ 
LHCb ~\cite{LHCBMIX2}  &$3.568\pm0.066$        &$4.8\pm1.0$           &$0.055\pm0.049$         \\ 
Belle (this work)      &$3.53\pm0.13$          &$4.6\pm3.4$           &$0.09\pm0.22$           \\ 
\end{tabular}     
\end{center}
\end{ruledtabular}
\end{table}

As a check of our results in Table~\ref{TABLE:MIXRESULTS}, we repeat
the analysis in two independent sub-samples. One corresponds to an
integrated luminosity of 400 fb$^{-1}$ (the ``old sample'') that is
used in our previous publication~\cite{OLDWS} with a different method
than used here. The other is the rest of our full data sample,
corresponding to an integrated luminosity of 576 fb$^{-1}$ (the ``new
sample''). These two independent sub-samples are fed through this
analysis separately. The results from the old and new samples (with
statistical uncertainty only) are $(R_D, y', {x'}^2)=(3.65\pm0.22,
-0.2\pm5.4, 0.36\pm0.32)\times 10^{-3}$ and $(3.45\pm0.17, 7.6\pm4.4,
-0.09\pm0.30)\times 10^{-3}$, respectively, which are compatible with
the results from the full data sample. Furthermore, the results of
this analysis using the old sample are consistent with our previous
publication~\cite{OLDWS}, which is superseded by the results of this
analysis.

\begin{figure}[htbp]
  \includegraphics[height=0.35\textwidth,width=0.47\textwidth]{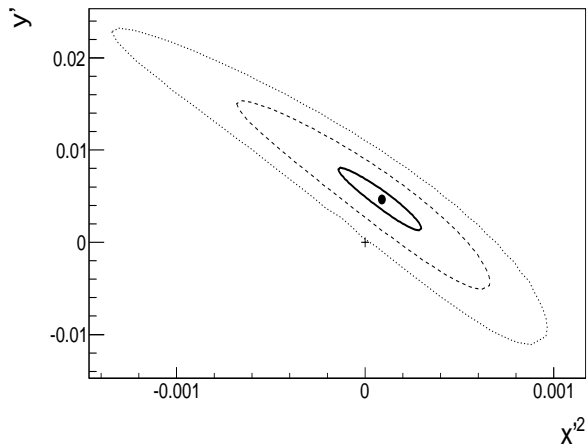}
  \caption{Best-fit point and contours in the $({x'}^2,y')$ plane. The
    solid, dashed, and dotted lines, respectively, correspond to 1, 3,
    and 5 standard Gaussian deviations from the best fit. The cross is
    the no-mixing point.}
\label{FIG:XYCONT}
\end{figure}

The $\chi^2$ difference between the ``no-mixing'' and ``mixing''
hypotheses, $\Delta\chi^2=\chi^2_{\rm no-mixing}-\chi^2_{\rm mixing}$,
is 29.3 for two degrees of freedom, corresponding to a probability of
4.3$\times10^{-7}$; this implies the no-mixing hypothesis is excluded
at the 5.1 standard deviation level. Thus, we observe $D^0-\bar{D}^0$
mixing for the first time in an $e^+e^-$ collision experiment. We also
show this in Fig.~\ref{FIG:XYCONT} with the 1$\sigma$, 3$\sigma$, and
5$\sigma$ contours around the best fit point in the $({x'}^2,y')$
plane.

In summary, we report the first observation of $D^0-\bar{D}^0$ mixing
in $e^+e^-$ collisions by measuring the time-dependent ratios of the
WS to RS decay rates, providing ${x'}^2=(0.09\pm0.22)\times 10^{-3}$,
$y'=(4.6\pm3.4)\times 10^{-3}$, and $R_D=(3.53\pm0.13)\times
10^{-3}$. Our results agree well with those from hadron collider
experiments~\cite{CDFMIX_PRE, LHCBMIX2} performed in very different
experimental conditions.

We thank the KEKB group for excellent operation of the accelerator;
the KEK cryogenics group for efficient solenoid operations; and the
KEK computer group, the NII, and PNNL/EMSL for valuable computing and
SINET4 network support. We acknowledge support from MEXT, JSPS and
Nagoya's TLPRC (Japan); ARC and DIISR (Australia); FWF (Austria); NSFC
(China); MSMT (Czechia); CZF, DFG, and VS (Germany); DST (India); INFN
(Italy); MOE, MSIP, NRF, GSDC of KISTI, BK21Plus, and WCU (Korea);
MNiSW and NCN (Poland); MES and RFAAE (Russia); ARRS (Slovenia);
IKERBASQUE and UPV/EHU (Spain); SNSF (Switzerland); NSC and MOE
(Taiwan); and DOE and NSF (USA). B.~R.~Ko acknowledges support by NRF
Grant No. 2010-0021279, and E.~Won by NRF Grant No. 2010-0021174.

\end{document}